\documentclass[]{aa}
\usepackage{amsmath}
\usepackage{amssymb}
\usepackage{natbib}
\usepackage{graphicx}
\usepackage{txfonts}
\usepackage[english]{babel}

\begin{document}

\title{Spectroscopic diagnostic for the mineralogy of large dust grains}
\author{M. Min \and C. Dominik \and L.~B.~F.~M. Waters}
\institute{Astronomical institute Anton Pannekoek, University of Amsterdam, Kruislaan 403, 1098 SJ  Amsterdam, The Netherlands\\
\email{mmin@science.uva.nl}}
\date{Submitted to A\&A Letters}

\abstract{We examine the thermal infrared spectra of large dust grains of
different chemical composition and mineralogy. Strong resonances
in the optical properties result in detectable spectral structure
even when the grain is much larger than the wavelength at which it
radiates. We apply this to the thermal infrared spectra of compact
amorphous and crystalline silicates. The weak resonances of
amorphous silicates at 9.7 and 18 $\mu$m virtually disappear for
grains larger than about 10 $\mu$m. In contrast, the strong
resonances of crystalline silicates produce \emph{emission dips}
in the infrared spectra of large grains; these emission dips are
shifted in wavelength compared to the \emph{emission peaks}
commonly seen in small crystalline silicate grains. We discuss the
effect of a fluffy or compact grain structure on the infrared
emission spectra of large grains, and apply our theory to the dust
shell surrounding Vega.} \maketitle

\section{Introduction}

Infrared spectroscopy is an invaluable tool for the study of the
structure and composition of interstellar and circumstellar dust.  In
the infrared, functional groups in a dust grain lead to spectral features in
the absorption cross section of a grain.  These features can be detected
in absorption against a strong infrared background source, or as 
emission features if the grains are warm enough to emit thermally in this 
wavelength region. The precise shapes and positions of such features contain 
information about grain size, shape and detailed chemical composition 
\citep{BohrenHuffman}.

It is generally assumed that dust grains need to be small
in order to produce spectral structure.  When grains are large, each
individual grain becomes optically thick at infrared
wavelength, and the features in the absorption cross section are
expected to weaken with increasing grain size and eventually to disappear
completely.  In most environments in interstellar space, small dust
grains dominate the grain surface available and therefore also
dominate the total absorption cross section.  There are, however,
environments where small grains are heavily depleted.  In particular
in gas-poor circumstellar disks, small dust grains are removed by
radiation pressure and larger grains start to dominate the interaction
with radiation.  For example, the zodiacal dust in the solar system is
dominated by large grains since grains smaller than about 1 micron are
quickly lost \citep{Burns1979}.  This effect becomes
even stronger in debris disks around A-type main sequence stars like
Vega, where the increased luminosity shifts the typical blowout size
to about 10 micron \citep{Artymowicz}.  Even stronger
effects may exist in disks around (post)-AGB stars, where already
grains of mm size will be lost unless the disk is optically thick and
gas-rich \citep{Dominik2003a}. Another application could be the emission spectra of asteroids of which the regolith is depleted from small dust grains \citep[and references therein]{Dollfus}.

We therefore examine in this paper the optical properties of large
dust grains.  In section \ref{sec:infr-emiss-spectra} we show that the
strong resonances observed in crystalline materials lead to observable
structure in the infrared spectra of large dust grains.
In section \ref{sec:spectrum-vega} we apply these results to the dust
shell around Vega.  In sections \ref{sec:discussion} and
\ref{sec:conclusions} we discuss and summarize our results.

\section{Infrared emission spectra}
\label{sec:infr-emiss-spectra}

\begin{figure*}[!t]
\resizebox{\hsize}{!}{\includegraphics{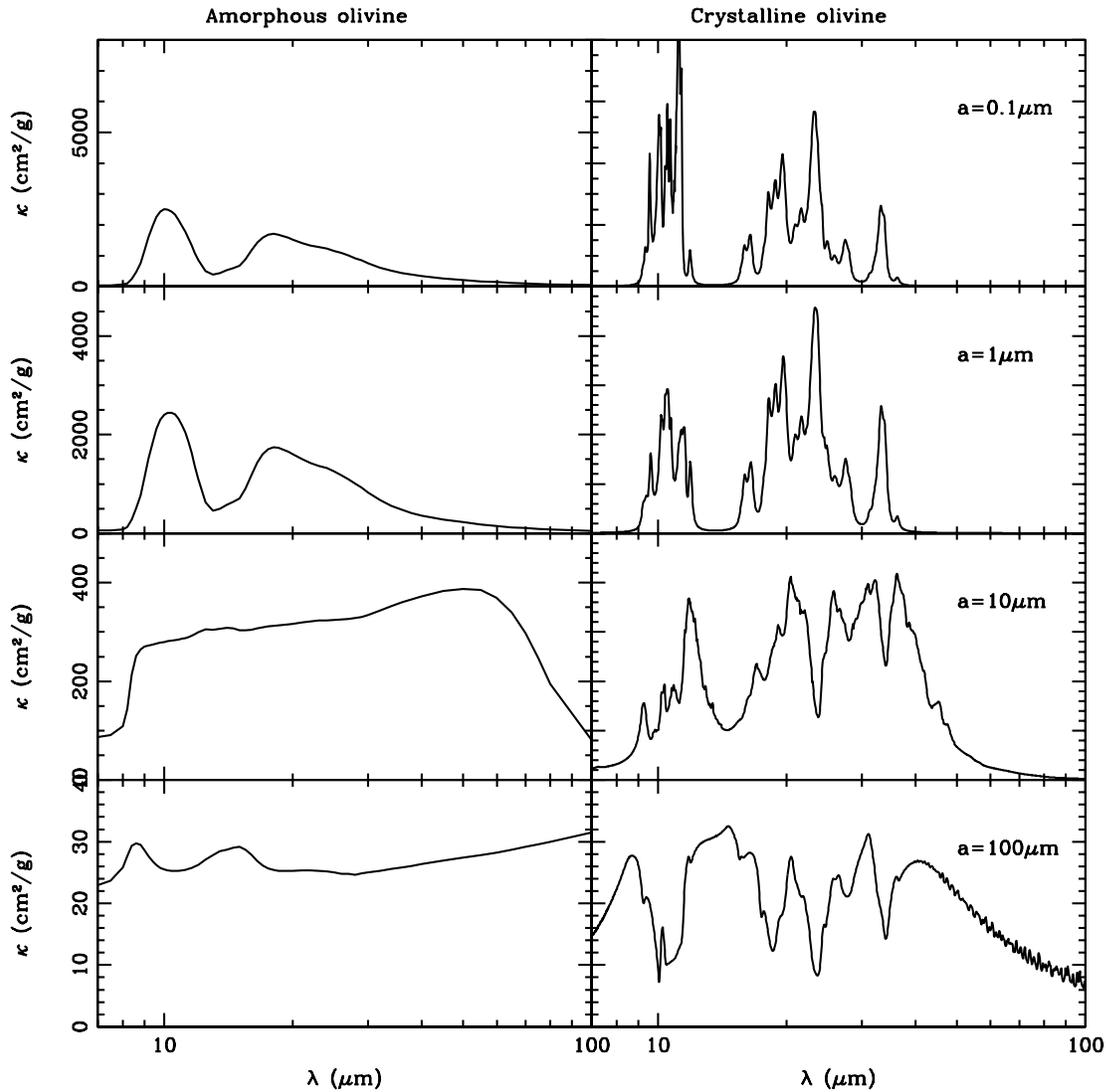}}
\caption{The emission cross sections per unit mass, $\kappa$, for
  various sizes of amorphous olivine grains (left panels) and crystalline 
  forsterite grains (right panels) as functions of the wavelength.
  The emissivities were calculated using a uniform distribution of
  spheroids (UDS). The refractive index data are taken from
  \citet{Servoin}. Note the different y-scales for the different grain
  sizes.}
\label{fig:kappas}
\end{figure*}

In order to calculate the thermal emission spectrum of a dust
shell we need to know the emission cross sections of the dust
grains. With grain size, shape and chemical composition known,
the emission cross section per unit mass as a function of wavelength
can be calculated from the wavelength dependent bulk refractive index.
In grains much smaller than the wavelength of radiation, a
resonance in the bulk refractive index causes a so-called surface mode
\citep{BohrenHuffman}. These surface modes are strong enhancements of
the absorption cross section and are detected as emission features in
the thermal emission spectra of these grains. For larger grains, the
surface modes disappear and it is generally assumed that the thermal
emission from large particles will be a smooth blackbody without a
characteristic signature of the mineralogy. However, this assumption
only holds if the refractive index is close to unity over the entire
wavelength range. For the resonances in crystalline silicates, both
the real and imaginary part of the refractive index can reach
values up to $\sim 9$. For large particles with these values of the
refractive index, most radiation will be reflected off the surface of
the particle and not penetrate the grain. This leads to a significant reduction 
of the grain absorption efficiency at these wavelengths. Therefore, we also 
expect to find a corresponding decrease of the emission efficiency, at 
wavelengths where there is a resonance in the bulk refractive index.  This is 
illustrated in Fig.~\ref{fig:kappas} where the emission cross sections per unit 
mass is plotted for grains composed of amorphous olivine and crystalline forsterite for different grain sizes. Moving from small to large grains, the spectral structure in the spectrum of the amorphous olivine virtually disappears. The resonances in the spectrum of the forsterite grains change from emission enhancements to emission dips. These dips will be detectable in the thermal emission spectra from large compact particles. Generally the features at shorter wavelengths and the strongest features are the first to disappear and change into emission dips. Most dips in the spectra of the large grains are slightly red-shifted compared to the corresponding emission enhancements in the small grain spectra. When the grain size is increased even more, the overall shape of the spectrum will flatten but the spectral structure remains visible at about the same contrast level.

\begin{figure*}[!tb]
\resizebox{\hsize}{!}{\includegraphics{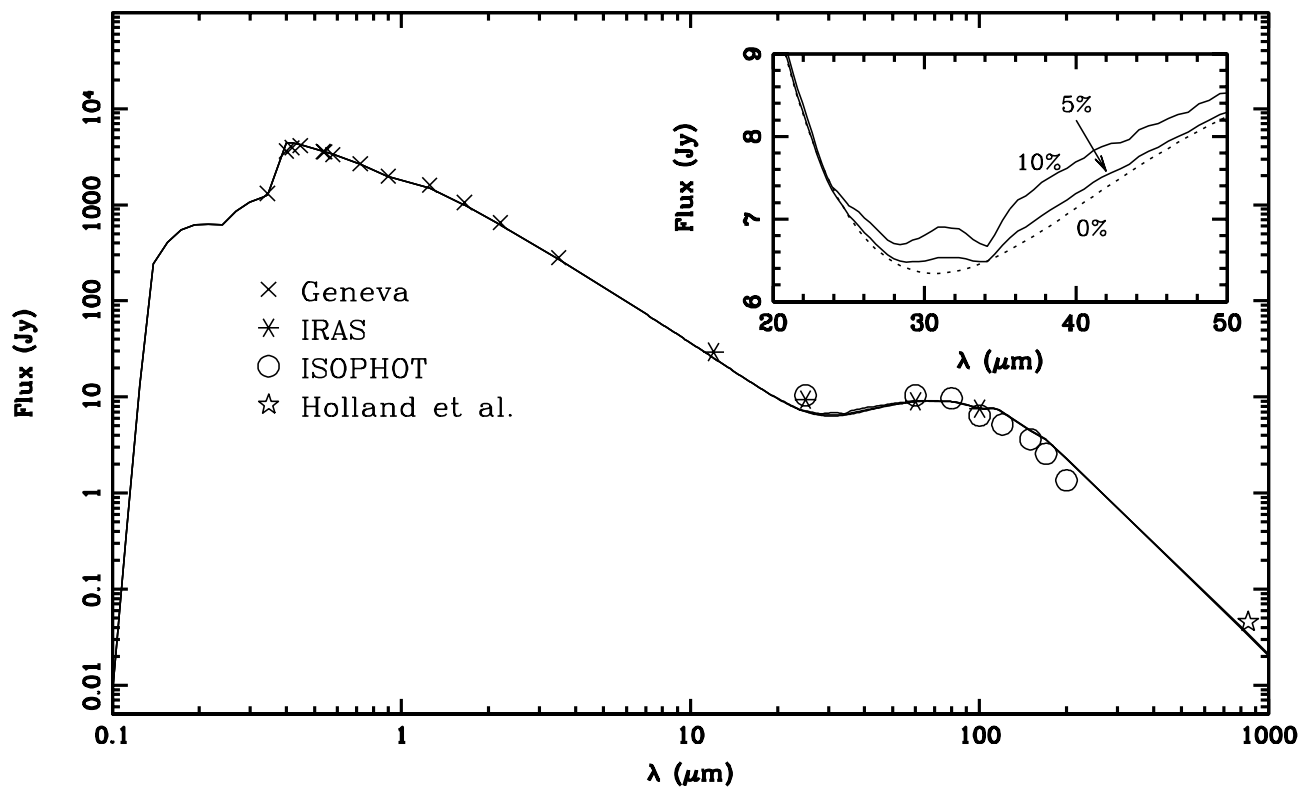}}
\caption{Model infrared spectra for Vega. The parameters for the model are given in the text. The IRAS data are taken from \cite{Walker1988}, the ISOPHOT data are from \cite{Heinrichsen1998} and the submm flux is taken from \cite{Holland}. The labels with the curves in the inset are the abundances of forsterite for the different models.}
\label{fig:spectrum}
\end{figure*}

In order to calculate the emission cross sections, a particle shape
has to be assumed.  We used a uniform distribution of spheroids which
is in very good agreement with laboratory measurements of the mass
absorption coefficients of small crystalline silicate grains
\citep{Min2003b}.  The cross sections were calculated using the
method described in \citet{Min2003a}.  For the large particles, the
emission efficiencies are the same for all convex shaped particles.
Calculations for homogeneous spheres using Mie theory \citep{Mie} will
therefore give the same emission spectrum in the large particle limit.

It is important to note that for large compact grains the emission cross 
section per unit mass is very low compared to small grains - this can be seen 
from the different scales used in fig.\ref{fig:kappas}. Also, the emission dips 
created in large grains are weaker than the emission peaks due to small grains. 
In most spectra, the emission by small grains, which is much more efficient, 
will be dominating.  The spectral signature of the large compact dust grains 
should therefore be observable only in environments where small grains are 
heavily depleted.

\section{The spectrum of Vega}
\label{sec:spectrum-vega}

An efficient mechanism to remove small grains from a population of
grains is radiation pressure on grains in orbit around a star.  Due to
the efficient absorption, the radiative acceleration on small grains
will easily exceed the gravitational force exerted by the stars.  If
the grains are orbiting in a gas-poor, optically thin circumstellar
shell, they will be removed on a Kepler timescale.  Such conditions
are given in the debris disks around main-sequence stars.

To demonstrate the effects of large compact crystalline grains, we
calculated the spectrum of Vega using a dust shell in a distance range
of 80-120 AU from the star.  The stellar spectrum is taken from a
Kurucz model \citep{Kurucz} with $T_{\mathrm{eff}}=9500$K and
$\log g=4$.  For the size distribution of the dust grains we used a
powerlaw with sizes between 15 and 100 $\mu$m and a powerlaw index of
$-3.5$, typical for a collisional cascade \citep{Tanaka1996}. From the
collisional model of \citet{Dominik2003} and the calculated radiation
pressure we estimated the fraction of small particles to be $\lesssim
0.1$\% of the total mass. Calculations show that this mass fraction
does not produce significant emission to be visible in the spectrum.
The dust is assumed to be a mixture of 20\% amorphous carbon and 80\%
olivine by mass. We calculated various models with different mass
fraction of crystalline and amorphous olivine. The dust grains are
assumed to be compact particles, the optical properties are calculated
using homogeneous spheres.  The dust temperatures are computed
self-consistently from the local radiation field. The total dust mass used in order to fit the photometry points is $\sim 1/10$ of a lunar mass which is in agreement with the estimate by \citet{Dent}.

The model spectrum is shown in Fig.~\ref{fig:spectrum}, which runs
reasonably well through the observed photometry points. 
It should be stressed here that the presented spectrum is an example to 
demonstrate the effect of large grains. We do not claim to provide the 
best fit to the available data points. 
The inset in Fig.~\ref{fig:spectrum} shows an enlargement of the 20 to 50 $\mu$m
region.  A prominent dip in the spectrum at 34 $\mu$m is caused by
large compact forsterite grains.  Even with only 5\% of the mass
present as compact forsterite grains, the dip would most likely be
detectable by SIRTF.

\section{Discussion}
\label{sec:discussion}

\subsection{Grain composition}

In section \ref{sec:spectrum-vega} we have shown that a fraction of about 5\% in compact forsterite grains would produce a detectable feature in the spectrum of Vega. It is important to note that this mass fraction should be present as seperate homogeneous grains. If the forsterite material is present as small inclusions inside an amorphous matrix, a significantly larger fraction of forsterite would be required to produce a spectral signature.

\subsection{Grain structure}

The structure of the dust grains is an important parameter that
contains information about its formation and processing history. As a
first rough classification we can distinguish between fluffy and
compact particles. The structure of a grain can be traced back to its
formation process. Dust grains that are grown by coagulation of small
(submicron sized) grains will show a very fluffy open structure \citep{Blum}, 
similar to the structure observed in interplanetary dust particles (IDP).  Such grains are also expected to result from the 
destruction of primitive planetary bodies such as small cometary nuclei.  
Compact (non-fluffy) large grains will result from the destruction of larger 
parent bodies which have been molten and possibly differentiated during the 
formation process, like large asteroids.

Measurements of the infrared transmission spectrum of a large
($a\approx20~\mu$m) fluffy IDP by \citet{Molster2003} still
shows the resonances characteristic for submicron grains. This
indicates that the infrared spectrum of fluffy particles will still be
dominated by the spectral signature of the constituents of the
aggregate. In other words, we expect to see the small particle
crystalline silicate emission features in the emission spectra of
these dust grains. For large compact particles we expect to see a
spectral signature as shown for 100 $\mu$m sized particles in
Fig.~\ref{fig:kappas}.

\section{Conclusions}
\label{sec:conclusions}

We have developed a new diagnostic tool for determining the
mineralogy of large compact dust grains.  The thermal emission spectra
of large compact crystalline silicate dust grains exhibit
characteristic features that can be detected with high signal to noise
spectroscopy.  These features are \emph{emission dips} at positions
that can be \emph{slightly red shifted} from the corresponding
emission peaks observed for small grains.  As an example, a model
calculation of the infrared emission from the dust shell around Vega
shows that these features should be detectable if the dust grains are
compact and contain at least 5\% crystalline forsterite in a seperate grain 
component.  Since there is a clear spectroscopic difference between fluffy and 
compact dust grains, the spectral signature of crystalline silicates can be 
used to obtain information on the structure, mineralogy and formation history 
of the dust grains. Other applications of this diagnostic can be the zodiacal 
dust or asteroidal regolith.

Our model calculations show that if large compact crystalline
silicates are present even with modest abundances, this should
be detectable as a depression near 34 $\mu$m. If the grains are
fluffy, an emission band should be present at a slightly shorter
wavelength. Such spectral structure may be measured
using the IRS spectrograph on board of SIRTF.

\begin{acknowledgements}
We would like to thank J.~W. Hovenier for valuable discussions. Comments on an earlier version of this manuscript by F.~J. Molster are gratefully acknowledged.
\end{acknowledgements}


\begin{thebibliography}{18}
\expandafter\ifx\csname natexlab\endcsname\relax\def\natexlab#1{#1}\fi

\bibitem[{Artymowicz(1988)}]{Artymowicz}
Artymowicz, P. 1988, ApJ, 335, L79

\bibitem[{Blum {et~al.}(2000)Blum, Wurm, Kempf, Poppe, Klahr, Kozasa, Rott,
  Henning, Dorschner, Schr{\"a}pler, Keller, Markiewicz, Mann, Gustafson,
  Giovane, Neuhaus, Fechtig, Gr{\"u}n, Feuerbacher, Kochan, Ratke, Goresy,
  Morfill, Weidenschilling, Schwehm, Metzler, \& Ip}]{Blum}
Blum, J., Wurm, G., Kempf, S., {et~al.} 2000, \prl, 85, 2426

\bibitem[{Bohren \& Huffman(1983)}]{BohrenHuffman}
Bohren, C.~F. \& Huffman, D.~R. 1983, Absorption and Scattering of Light by
  Small Particles (New York: Wiley)

\bibitem[{Burns {et~al.}(1979)Burns, Lamy, \& Soter}]{Burns1979}
Burns, J.~A., Lamy, P.~L., \& Soter, S. 1979, Icarus, 40, 1

\bibitem[{Dent {et~al.}(2000)Dent, Walker, Holland, \& Greaves}]{Dent}
Dent, W. R.~F., Walker, H.~J., Holland, W.~S., \& Greaves, J.~S. 2000, \mnras,
  314, 702

\bibitem[{Dollfus(1990)}]{Dollfus}
Dollfus, A. 1990, in Asteroids, comets, meteors III, ed. C.~I. Lagerkvist,
  H.~Rickman, \& B.~A. Lindblad, 49

\bibitem[{Dominik \& Decin(2003)}]{Dominik2003}
Dominik, C. \& Decin, G. 2003, ApJ, in press

\bibitem[{Dominik {et~al.}(2003)Dominik, Dullemond, Cami, \& van
  Winckel}]{Dominik2003a}
Dominik, C., Dullemond, C.~P., Cami, J., \& van Winckel, H. 2003, A\&A, 397,
  595

\bibitem[{Heinrichsen {et~al.}(1998)Heinrichsen, Walker, \&
  Klaas}]{Heinrichsen1998}
Heinrichsen, I., Walker, H.~J., \& Klaas, U. 1998, \mnras, 293, L78

\bibitem[{Holland {et~al.}(1998)Holland, Greaves, Zuckerman, Webb, McCarthy,
  Coulson, Walther, Dent, Gear, \& Robson}]{Holland}
Holland, W.~S., Greaves, J.~S., Zuckerman, B., {et~al.} 1998, Nature, 392, 788

\bibitem[{Kurucz(1993)}]{Kurucz}
Kurucz, R. 1993, Model Atmospheres, CD-ROM Nos. 1-18, Cambridge, Mass.:
  Smithsonian Astrophysical Observatory

\bibitem[{Mie(1908)}]{Mie}
Mie, G. 1908, Ann Phys., 25, 377

\bibitem[{Min {et~al.}(2003{\natexlab{a}})Min, Hovenier, \&
  de~Koter}]{Min2003a}
Min, M., Hovenier, J.~W., \& de~Koter, A. 2003{\natexlab{a}}, \jqsrt, 79--80,
  939

\bibitem[{Min {et~al.}(2003{\natexlab{b}})Min, Hovenier, \&
  de~Koter}]{Min2003b}
---. 2003{\natexlab{b}}, A\&A, 404, 35

\bibitem[{Molster {et~al.}(2003)Molster, Demyk, D'Hendecourt, \&
  Bradley}]{Molster2003}
Molster, F.~J., Demyk, A., D'Hendecourt, L., \& Bradley, J.~P. 2003, 34th
  Annual Lunar and Planetary Science Conference, abstract nr. 1148

\bibitem[{Servoin \& Piriou(1973)}]{Servoin}
Servoin, J.~L. \& Piriou, B. 1973, Phys. Stat. Sol. (b), 55, 677

\bibitem[{Tanaka {et~al.}(1996)Tanaka, Inaba, \& Nakazawa}]{Tanaka1996}
Tanaka, H., Inaba, S., \& Nakazawa, K. 1996, Icarus, 123, 450

\bibitem[{Walker \& Wolstencroft(1988)}]{Walker1988}
Walker, H.~J. \& Wolstencroft, R.~D. 1988, \pasp, 100, 1509

\end{thebibliography}

\end{document}